\documentclass[11pt]{article}
\usepackage{amsmath,epsf,amssymb,latexsym,enumerate,cite,shadow,array,color}
\usepackage{slashed}
\usepackage{graphicx,wasysym}
 \pdfoutput=1

\DeclareFontFamily{U}{rsf}{} \DeclareFontShape{U}{rsf}{m}{n}{
  <5> <6> rsfs5 <7> <8> <9> rsfs7 <10-> rsfs10}{}
\DeclareMathAlphabet\Scr{U}{rsf}{m}{n} \makeatletter
\@addtoreset{equation}{section} \makeatother

\def\be{\begin{equation}}
\def\ee{\end{equation}}
\def\ba{\begin{array}}
\def\ea{\end{array}}

\newcommand{\bea}{\begin{eqnarray}}
\newcommand{\eea}{\end{eqnarray}}

\def\K{K{\"a}hler}
\def\vp{{\varphi}}

\newcommand{\ft}[2]{{\textstyle\frac{#1}{#2}}}

\def\rme{{\rm e}}

\usepackage{ifpdf}
\ifx\pdfoutput\undefined
   \pdffalse
\else
   \pdfoutput=1
   \pdftrue
  \usepackage[pdftex]{hyperref}
\newcommand{\rf}[1]{(\ref{#1})}

\begin{document}

\begin{titlepage}

\

\

\

\

\begin{center}
{\huge \textbf{Multi-field Conformal Cosmological Attractors
\vskip 0.8cm }}

\

\

{\bf Renata Kallosh} and {\bf Andrei Linde}  

\vskip 0.3cm

{\sl Department of Physics and SITP, Stanford University, Stanford, California
94305 USA}\\
\end{center}
\vskip 1 cm

\

\begin{abstract}

We describe a broad class of multi-field inflationary models with spontaneously broken conformal invariance. It generalizes the recently discovered class of cosmological attractors with a single inflaton field \cite{Kallosh:2013hoa}. In the new multi-field theories, just as in the single-field models of  \cite{Kallosh:2013hoa}, the moduli space has a boundary (\K\  cone)  in terms of the original homogeneous conformal variables.  Upon spontaneous breaking of the conformal invariance and switching to the Einstein frame, this boundary moves to infinity in terms of the canonically normalized inflaton field. This results in the exponential stretching and flattening of scalar potentials in the vicinity of the boundary of the moduli space, which makes even very steep potentials perfectly suitable for the slow-roll inflation. These theories, just like their single-field versions, typically lead to inflationary perturbations with $n_{s} =1-2/N$ and $r = 12/N^{2}$, where $N$ is the number of e-foldings. 
 \end{abstract}

\vspace{24pt}
\end{titlepage}




\section{Introduction}

In  our recent paper \cite{Kallosh:2013hoa} we have found a broad class of models with spontaneously broken conformal or superconformal invariance which allow inflation even in the theories with very steep potentials in terms of the original conformal variables. This class of theories have universal observational predictions $1 -n_{s} =2/N$ and $r = 12/N^{2}$, which are very stable with respect to strong modifications of the inflationary potential in these models, and are in perfect agreement with the recent observational data from WMAP9 \cite{Hinshaw:2012aka} and Planck 2013 \cite{Ade:2013rta}. The main reason of this universality is the exponential stretching of the boundary of the moduli space (the \K\ cone) upon switching to the canonically normalized fields in the Einstein frame. 

These predictions coincide with predictions of the theory $R +\alpha R^{2}$ \cite{Starobinsky:1980te,Kofman:1985aw}, which can be naturally embedded into this new class of models \cite{Kallosh:2013hoa,Kallosh:2013lkr}. The same observational predictions were made in the context of chaotic inflation in the theory $\lambda\phi^{4}$ \cite{Linde:1983gd} with non-minimal coupling to gravity ${\xi\over 2}\phi^{2} R$ with $\xi > 0$ \cite{Salopek:1988qh,Sha-1,Kaiser:2013sna},  in its supersymmetric extensions in \cite{Einhorn:2009bh,Ferrara:2010yw,Lee:2010hj,Ferrara:2010in,Kallosh:2013pby,Ferrara:2013kca}
and in a certain limit of the model with the Higgs potential with $\xi < 0$ \cite{Linde:2011nh}. Recently another set of models with similar predictions was identified in \cite{Ellis:2013xoa,Buchmuller:2013zfa,Farakos:2013cqa,Ellis:2013nxa,Ferrara:2013rsa,Kallosh:2013maa}, and some earlier works in the related area, such as \cite{Cecotti:1987sa,Cecotti:1987qe}, have been rediscovered and reinforced \cite{Kallosh:2013lkr,Farakos:2013cqa,Ferrara:2013rsa,Ferrara:2013kca}. 

In the models studied in \cite{Kallosh:2013hoa} we concentrated on the theories with a single inflaton field, other scalar fields being strongly stabilized. In this paper we are going to generalize these results to models of many scalars with spontaneously broken conformal or superconformal invariance.  In section 2 of this paper  we will give a brief summary of the results of Ref. \cite{Kallosh:2013hoa}. In Section 3 we will generalize these results for the theories with two different inflaton fields with simplest potentials, resembling the T-Model potentials introduced in  \cite{Kallosh:2013hoa}.  In Section 4 we will discuss inflation in the theories with more general potentials.  
 
 \section{T-Model and single-field inflationary attractors}\label{single}
 
In order to explain the basic idea of our approach, we will remember the main features of the single inflaton field models proposed in \cite{Kallosh:2013hoa}. We will begin with a toy model with the following Lagrangian:
\begin{equation}
\mathcal{L} = \sqrt{-{g}}\left[{1\over 2}\partial_{\mu}\chi \partial^{\mu}\chi  +{ \chi^2\over 12}  R({g})- {1\over 2}\partial_{\mu} \phi\partial^{\mu} \phi   -{\phi^2\over 12}  R({g}) -{\lambda\over 36} (\phi^{2}-\chi^{2})^{2}\right]\,.
\label{toy}
\end{equation}
This
theory is locally conformal invariant under the following
transformations: 
\be \tilde g_{\mu\nu} = \rme^{-2\sigma(x)} g_{\mu\nu}\,
,\qquad \tilde \chi =  \rme^{\sigma(x)} \chi\, ,\qquad \tilde \phi =  \rme^{\sigma(x)}
\phi\ . \label{conf}\ee 
In addition, it has a global $SO(1,1)$ symmetry with respect to a boost between these two fields, preserving the value of $\chi^2-\phi^2$, which resembles Lorentz symmetry of special theory of relativity.

The field $\chi(x)$ is the conformon.  It can be removed from the theory by fixing the gauge symmetry
(\ref{conf}), for example by taking a gauge $\chi =\sqrt 6$. This gauge fixing can be interpreted as a spontaneous breaking of conformal invariance due to existence of a classical field $\chi =\sqrt 6$. This will bring the theory to the Jordan frame, from which one may further proceed to the Einstein frame.

 However, one can obtain the same results in a much easier way by using the gauge $\chi^2-\phi^2=6$ and resolving this constraint in terms of the  canonically normalized field $\varphi$: 
$\chi=\sqrt 6 \cosh  {\varphi\over \sqrt 6}$, $ \phi= \sqrt 6 \sinh {\varphi\over \sqrt 6} $.
Our action \rf{toy} becomes
\begin{equation}\label{chaotmodel1}
L = \sqrt{-g} \left[  \frac{1}{2}R - \frac{1}{2}\partial_\mu \varphi \partial^{\mu} \varphi -   \lambda \right].
\end{equation}
Thus our original theory is equivalent to a theory of gravity, a free massless canonically normalized field $\varphi$, and a cosmological constant $\lambda$  \cite{Kallosh:2013hoa}.

The main reason for doing this exercise was to show that the somewhat unusual term $(\phi^{2}-\chi^{2})^{2}$, or similar terms which will appear later in our paper, are essentially the placeholders to what will eventually look like a cosmological constant in the Einstein frame. The theories to be studied below are based on the idea that one can develop an interesting class of inflationary models by modifying these placeholders, i.e. by locally deforming the would-be cosmological constant. In order to do it without violating the original conformal invariance of the theory one can multiply the term $(\phi^{2}-\chi^{2})^{2}/36$ by an arbitrary function $F(z)$, where the variable $z = {\phi/\chi}$ is the homogeneous variable which properly describes the shape of the function $F\left({\phi/\chi}\right)$ in a conformally invariant way:
\begin{equation}
\mathcal{L} = \sqrt{-{g}}\left[{1\over 2}\partial_{\mu}\chi \partial^{\mu}\chi  +{ \chi^2\over 12}  R({g})- {1\over 2}\partial_{\mu} \phi\partial^{\mu} \phi   -{\phi^2\over 12}  R({g}) -{1\over 36} F\left({\phi/\chi}\right)(\phi^{2}-\chi^{2})^{2}\right]\,.
\label{chaotic}
\end{equation}

Using the gauge  $\chi^2-\phi^2=6$ immediately transforms the theory to the following equivalent form:
\begin{equation}\label{chaotmodel}
L = \sqrt{-g} \left[  \frac{1}{2}R - \frac{1}{2}\partial_\mu \varphi \partial^{\mu} \varphi -   F(\tanh{\varphi\over \sqrt 6}) \right].
\end{equation}
If, as we assume, the function $F(\tanh{\varphi\over \sqrt 6})$ is non-singular, then asymptotically $\tanh\varphi\rightarrow \pm 1$ and therefore $F(\tanh{\varphi\over \sqrt 6})\rightarrow \rm const$. Therefore the system in the large $\varphi$ limit evolves asymptotically  towards its critical point  where the $SO(1,1)$ symmetry is restored.

It is useful to present an alternative, more conventional derivation of the same result, using the gauge $\chi(x) = \sqrt{6}$ instead of  the gauge $\chi^2-\phi^2=6$. The full
Lagrangian in the Jordan frame becomes
\begin{equation}
\mathcal{L}_{\rm total }= \sqrt{-{g_{J}}}\,\left[{  R({g_{J}})\over 2}\left(1-{ \phi^2\over 6}\right)-  {1\over 2}\partial_{\mu} \phi \partial^{\mu} \phi    -F\left(\phi/\sqrt 6\right) \left({\phi^2\over 6}-1 \right)^{2}\right]\,.
\label{toy2j}
\end{equation}
Now one can represent the same theory in the Einstein frame, by changing the metric $g_J$ and $\phi$ to a conformally related metric $g_{E}^{\mu\nu} = (1- \phi^{2}/6)^{{-1}}  g_{J}^{\mu\nu}$ and a canonically normalized  field $\varphi$ related to the field $\phi$ as follows:
\be\label{field1j}
 \frac{d\varphi}{d\phi} = {1\over 1- {\phi^2/ 6}} \ .
\ee
In new variables, the Lagrangian, up to a total derivative, is given by
\begin{equation}\label{LE1}
L = \frac{1}{2} \sqrt{-g} \left[ R - \frac{1}{2}g^{\mu\nu} \partial_\mu \varphi \partial_\nu \varphi -  V(\phi(\varphi)) \right],
\end{equation}
where the potential in the Einstein frame is 
\begin{equation}\label{eframej}
V(\phi) = {1\over 36} F\left(\phi/\sqrt{6}\right)  {\left(\phi^2-6 \right)^{2}\over \left(1-{ \phi^2\over 6}\right)^{2}} =  F(\tanh{\varphi\over \sqrt 6}) = V(\vp) \ .
\end{equation}
This  brings the theory to the form   (\ref{chaotmodel}).

This approach allows us to explain the difference between our class of models and the more standard class of conformal models where the investigation begins with the Jordan frame with generic potentials $V_{J}(\phi)$ such as $\phi^{4}$. In such models, the transition to the Einstein frame results in singular potentials 
\be\label{jordan}
V_{E}(\phi) = {V_{J}(\phi)\over \left(1-{ \phi^2\over 6}\right)^{2}} 
\ee
Upon transitions to the canonical variables $\vp$, the singularity goes to infinity, but the potentials typically grow too fast at large $\vp$ to serve as  slow-roll inflationary potentials. That is why the models with $\xi = -1/6$, and with any other negative values of $\xi$, have not been very popular among those who wanted to construct viable inflationary models. 

However, if we approach it from the point of view of the conformally invariant theory with SO(1,1) symmetry (\ref{toy}), and then make its generalization (\ref{chaotic}), we automatically have the term $(\phi^{2}-\xi^{2})^{2}$ in $V_{J}(\phi)$, which exactly cancels the singularity present in (\ref{jordan}) upon the gauge fixing $\xi = \sqrt 6$.  The gauge choice $\chi^{2}-\phi^{2} = 6$, which we used earlier, provided a convenient shortcut to the same result. Once it is done, we have a large variety of models perfectly suited for inflation, with the cosmological predictions which are very stable with respect to the choice of the function $F\left({\phi/\chi}\right)$.

\begin{figure}[ht!]
\centering
\includegraphics[scale=1]{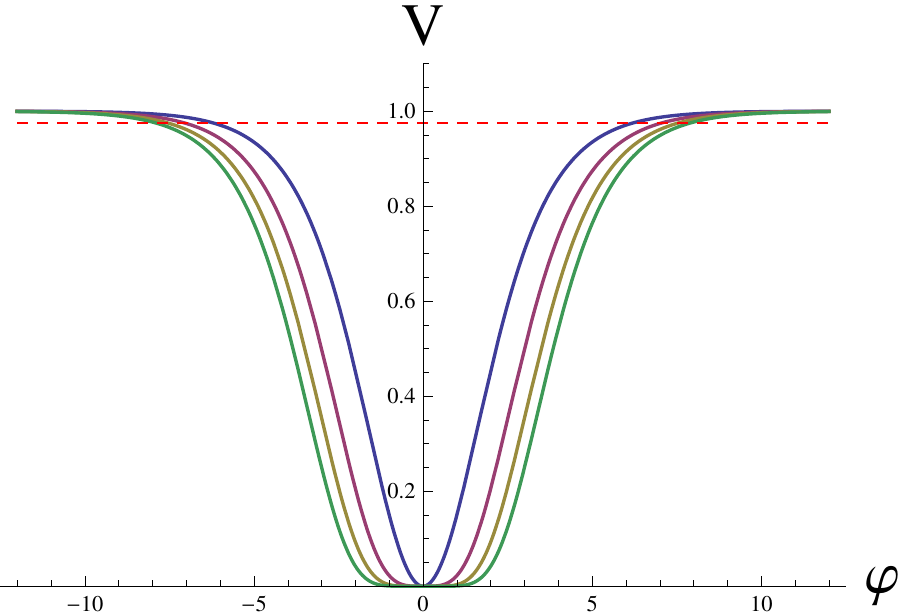}
\caption{\small Potentials for the T-Model inflation ${\tanh}^{2n}(\varphi/\sqrt6)$ for $n = 1,2,3,4$ (blue, red, brown and green,  corresponding to increasingly wider potentials). All of these models predict the same values $n_{s} =1-2/N$, $r = 12/N^{2}$ in the leading approximation in $1/N$. Here $N\sim 60$ is the number of e-foldings. The points where each of these potentials cross the red dashed line $V = 1-3/2N = 0.975$ correspond to the values of  $\vp$ where the perturbations are produced on scale corresponding to $N = 60$. For the simplest theory ${\tanh}^{2}(\varphi/\sqrt6)$ this point is at about $\varphi \sim 5.3$ in Planck units.}
\label{tmodelfig}
\end{figure}

As an example, one may consider the simplest set of functions $F\left({\phi/\chi}\right) = \lambda_{n} \left({\phi/\chi}\right)^{2n} = \lambda_{n} z^{2n}$. This is a generalization of the standard approach to chaotic inflation, where we originally used the simplest choice of functions $~\phi^{2n}$ \cite{Linde:1983gd}. Now we are making a similar choice, but in term of the homogeneous variables $z = \phi/\chi$, which preserve conformal invariance.
In this case one finds 
\begin{equation}\label{TModel}
V(\varphi) = \lambda_n\ {\tanh}^{2n}(\varphi/\sqrt6) .
\end{equation}
We called these models `T-Models'   because they originate from different powers of $\tanh(\varphi/\sqrt6) $,  and the potentials have the shape of the letter T, see Fig. \ref{tmodelfig}. Even though these potentials depend on $n$, observational predictions of these models do not depend on $n$.  Its basic representative $ \lambda_1\ {\tanh}^{2}(\varphi/\sqrt6)$ is the simplest version of the class of conformal chaotic inflation models proposed in \cite{Kallosh:2013hoa}. Rather unexpectedly, inflation is possible in a very broad class of models of this type, including models with very steep potentials $V(\phi)$ in terms of the original field variable $\phi$. Moreover, almost all such models have nearly identical observational consequences, in the leading order in $1/N$, thus belonging to the same universality class \cite{Kallosh:2013hoa}. Recently these conclusions have been confirmed by explicit calculations of $n_{s}$ and $r$ in the next-to-leading order in $1/N$ \cite{Roest:2013fha}.

\begin{figure}[ht!]
\centering
\includegraphics[scale=0.21]{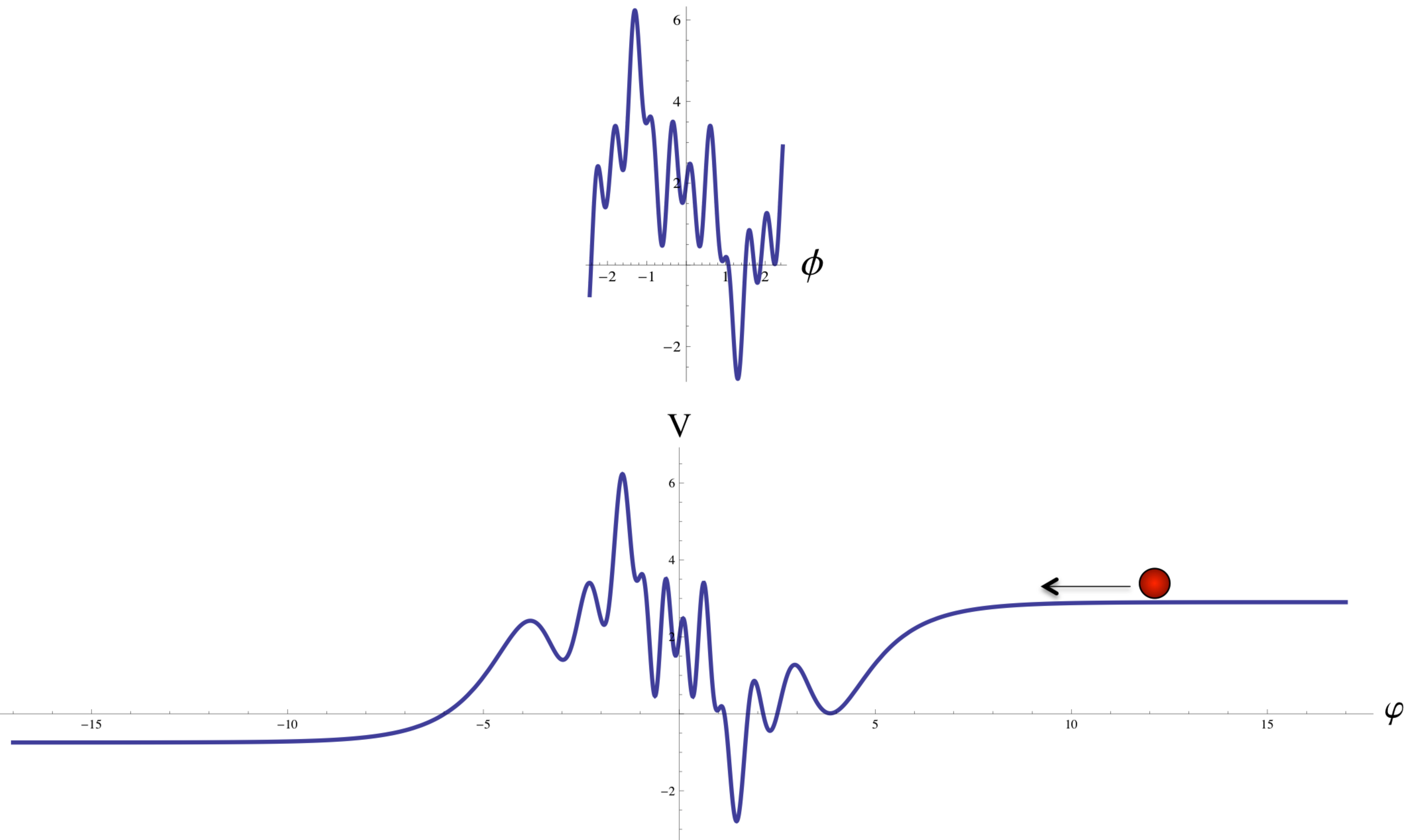}
\caption{\small Basic mechanism which leads to inflation in the theories with generic functions $F(z)$. The potential $V$ in the Einstein frame can have an arbitrary shape in terms of the original conformal variable $z$, which becomes $\phi/\sqrt 6$ in the gauge $\chi = \sqrt 6$; see e.g. the potential $V(\phi/\sqrt 6)$ in  the upper panel. If this potential is non-singular at the boundary of the moduli space $|z| = 1$ ($\phi= \sqrt 6$), it looks exponentially stretched and flat at large values of the canonically normalized field $\varphi$. This stretching  makes inflation very natural, and leads to universal observational predictions for a very broad class of such models \cite{Kallosh:2013hoa}. }
\label{stretch}
\end{figure}

To gain  intuitive understanding of these features of the new class of models, following  \cite{Kallosh:2013hoa}, let us consider an arbitrary non-singular potential $V(\phi) =F\left(\phi/\sqrt{6}\right)$, as shown in the upper panel of Fig. \ref{stretch}, and then plot the same potential in terms of the canonically normalized field $\varphi$ in the Einstein frame, shown in the lower panel of Fig. \ref{stretch}. As we see, the potential looks extremely flat upon switching to the canonically normalized field $\varphi$, which exponentially stretches and flattens the potential close to the boundary of the moduli space, sending this boundary to infinity, in terms of the field $\varphi$.\footnote{Other ways of flattening of the inflaton potentials were proposed in \cite{Dong:2010in}.} In the next sections we will generalize this mechanism for the multi-field model case.

\

\section{T-Model and multiple-field inflationary attractors}\label{dSsection}

Now we will consider a model with the following Lagrangian of the conformon field $\chi$  and  the fields $\phi_{i}$, where $i= 1,...,n$:
\begin{equation}
\mathcal{L} = \sqrt{-{g}}\left[{1\over 2}\left((\partial_{\mu}\chi)^{2} - (\partial_{\mu} \phi_{i})^{2}\right) +{ \chi^2-\phi_{i}^{2}\over 12}  R({g})  -{F\left({\phi_{i}\over\chi}\right)\over 36} (\chi^{2}-\phi_{i}^{2})^{2}\right]\,.
\label{toy2}
\end{equation}
 This theory is locally conformal invariant under the following
transformations: 
\be \tilde g_{\mu\nu} = \rme^{-2\sigma(x)} g_{\mu\nu}\,
,\qquad \tilde \chi =  \rme^{\sigma(x)} \chi\, ,\qquad \tilde \phi_{i} =  \rme^{\sigma(x)}
\phi_{i}\ . \label{conf2}\ee 
In addition, it has a global $SO(n,1)$ symmetry.

For simplicity, we will discuss here only two fields $\phi_{i}$, and represent them as follows:
\be
\phi_{1} = \rho \cos \theta, \qquad \phi_{2} = \rho \sin \theta \ .
\ee
The Lagrangian becomes
\begin{equation}
\mathcal{L} = \sqrt{-{g}}\left[{1\over 2}\left((\partial_{\mu}\chi)^{2} - (\partial_{\mu} \rho)^{2}- \rho^{2} (\partial_{\mu} \theta)^{2}\right) +{ \chi^2-\rho^{2}\over 12}  R({g})  -{F\bigl({\rho\over\chi},\,\theta\bigr)\over 36} (\chi^{2}-\rho^{2})^{2}\right]\,.
\label{spherical}
\end{equation}
As in the previous section, we will use the gauge $\chi^2-\rho^2=6$ and resolve this constraint in terms of the  canonically normalized field $\varphi$: 
$\chi=\sqrt 6 \cosh  {\varphi\over \sqrt 6}$, $ \rho= \sqrt 6 \sinh {\varphi\over \sqrt 6} $.
Our action \rf{toy2} becomes
\begin{equation}\label{chaotmodel1aa}
L = \sqrt{-g} \left[  \frac{1}{2}R - \frac{1}{2}(\partial_\mu \varphi)^{2} - {3}\sinh^{2} {\varphi\over \sqrt 6}\  (\partial_{\mu} \theta)^{2} -  F\bigl({\tanh{\varphi\over \sqrt 6}},\,\theta\bigr) \right].
\end{equation}
Thus our original theory describes the theory of gravity with a cosmological constant $\lambda$, and also two scalar fields: a canonically normalized field $\varphi$ and a non-canonically normalized field $\theta$, with the Einstein frame potential
\be
V = F\bigl({\tanh{\varphi\over \sqrt 6}},\,\theta\bigr).
\ee

\begin{figure}[ht!]
\centering
\includegraphics[scale=0.3]{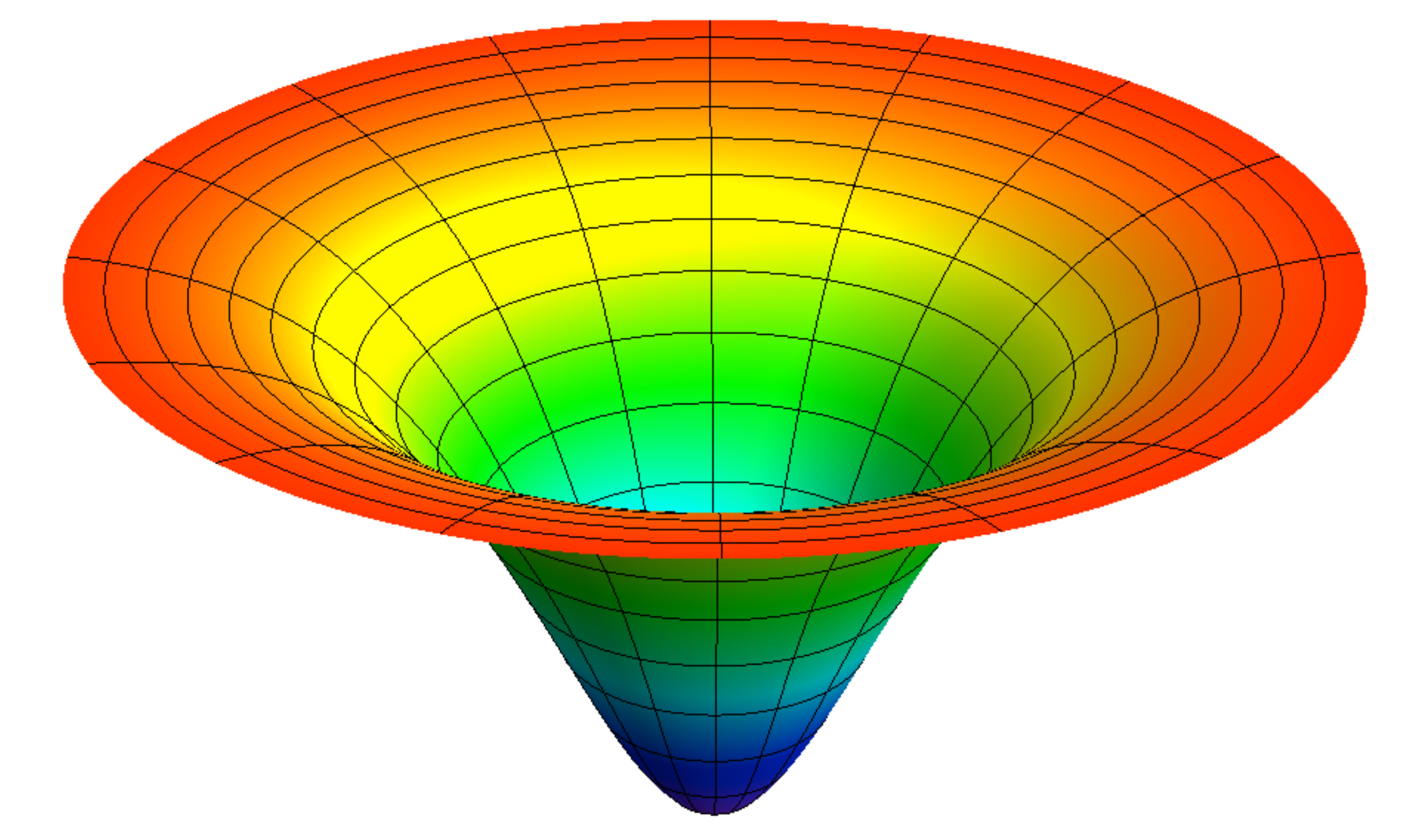}
\caption{\small T-Model for two light scalars described by (\ref{chaotmodel1aa}). The height of the potential is color-coded. This helps to instantly recognize asymptotically flat inflationary directions of the potential. The potential extends to infinitely large values of the canonically normalized field $\vp$ and becomes absolutely flat in the large $\vp$ limit. The figure shows only the part of the potential responsible for the last 60 e-foldings of inflation.}
\label{tmodelfig1}
\end{figure}

To explore the new set of possibilities, we will start with the multi-field T-Model potential. It originates from $F(z_{i}) = V_{0}\, z^{2}_{i}$ and yields
\be\label{2dt}
V =V_{0} \, {\tanh^{2}{\varphi\over \sqrt 6}}.
\ee
The potential does not depend on $\theta$. It is shown in Fig. \ref{tmodelfig1}, up to the level  $0.975\, V_{0}$, which corresponds to the value of $\vp = \vp_{60}\sim 6.2$, starting from which the universe inflates $e^{60}$ times while the field falls to the minimum of the potential. The potential extends to infinitely large values of the canonically normalized field $\vp$, but all or almost all that we need to know to describe formation of the observable part of the universe is shown in this figure.

This and other figures which we are going to show are the plots in the coordinates $(\vp,\theta)$. They can serve our intuition well by appropriately describing the curvature of the potential in the canonically normalized radial direction $\vp$, but they exponentially under-represent flatness of the potential in the angular direction. That is because the coordinates  $\chi$ and $\rho$ on the hyperboloid $\chi^2-\rho^2=6$ depend on $\vp$ exponentially. Therefore the canonically normalized length of the circle with the ``radius'' $\vp \sim 6.2$ shown in Fig. \ref{tmodelfig1} is not given by $2\pi \vp \sim 39$, as one could naively expect, but by  $2\pi \sqrt 6 \sinh {\varphi\over \sqrt 6} \sim 96$, see (\ref{chaotmodel1aa}).  For larger values of $\vp$, this difference becomes exponentially large, which leads to exponential suppression of the effective mass of the angular component of the field.

This is somewhat different from what happens in the multi-filed Higgs inflation with nonminimal positive coupling $\xi > 0$ studied in  \cite{Kaiser:2013sna}, where the function in front of $(\partial_{\mu} \theta)^{2}$ in the corresponding Lagrangian does not grow exponentially for large $\vp$. Instead of that, it grows very slowly, asymptotically approaching the constant $\xi^{{-1}}$. As a result, the effective mass of the angular component in the multi-scalar theories with $\xi > 0$ may become large at large $\vp$, whereas the mass of the inflaton field $\vp$ in the class of theories of the type of $\lambda\phi^{4}$ becomes exponentially small in the large $\vp$ limit. Therefore the field $\theta$ in such models with  $\xi > 0$ typically very rapidly rolls to the minimum of its potential, and the remaining cosmological evolution becomes dominated by the single-field evolution of the radial component of the field, i.e. by the inflaton field $\vp$  \cite{Kaiser:2013sna}.

The situation in our class of models with conformal coupling $\xi = -1/6 <0$ is more nuanced, but the final conclusion is quite similar: In most cases, the final stage of the inflationary evolution is determined by the single-field motion of the field $\vp$. 

Equations of motion for the fields $\vp$ and $\theta$ in this theory look as follows:
\be\label{s1}
\ddot\theta +\sqrt{2\over 3}\, {\dot\vp\, \dot\theta\over \tanh{\vp\over \sqrt{6}}} +3H\dot\theta = -{V_{\theta}\over {6}\sinh^{2} {\varphi\over \sqrt 6}} \ ,
\ee
\be\label{s2a}
\ddot\vp + 3H\dot \vp = \sqrt{3\over 2}\, \dot\theta^{2}\, \sinh{ \sqrt{2\over 3}\vp} - V_{\vp}\ .
\ee

We will study the evolution of the fields during inflation at asymptotically large $\vp$, where the terms ${V_{\theta}\over {6}\sinh^{2} {\varphi\over \sqrt 6}}$ and $V_{\vp}$ in the right hand sides of these equations are exponentially small. An investigation of a combined evolution of the fields $\vp$ and $\theta$ in this case shows that if the evolution begins at sufficiently large values of the field $\vp$ in the region with $V >0$, then after a short period of relaxation, the fields approach a slow roll regime where the kinetic energy of both fields is much smaller than $V_{0}$. Under this assumption, one can neglect the first two terms in equation (\ref{s1}) and the first term in equation (\ref{s2a}). Taking into account that  $\tanh{\vp\over \sqrt{6}} \approx 1$ and $\sinh{\varphi\over \sqrt 6} \approx {1\over 2}\, e^{\varphi/\sqrt 6}$ at $\vp \gg 1$, one can represent equations for the fields $\theta$ and $\vp$ in the slow roll approximation as follows:
\be
3H\dot\theta = -{2\over 3}\, e^{-\sqrt{2\over 3}\vp}\ V_{\theta} \ ,
\ee
\be
3H\dot \vp = {1\over 2}\sqrt{3\over 2}\ \dot\theta^{2}\, e^{\sqrt{2\over 3}\vp} - V_{\vp}\ .
\ee
where $H^{2} = V(\vp,\theta)/3$. Using the first of these two equations, the second one can be represented as
\be
3H\dot \vp = {1\over 9}\sqrt{2\over 3}\ {V^{2}_{\theta}\over V}\, e^{-\sqrt{2\over 3}\vp} - V_{\vp}\ .
\ee

\begin{figure}[ht!]
\centering
\includegraphics[scale=0.4]{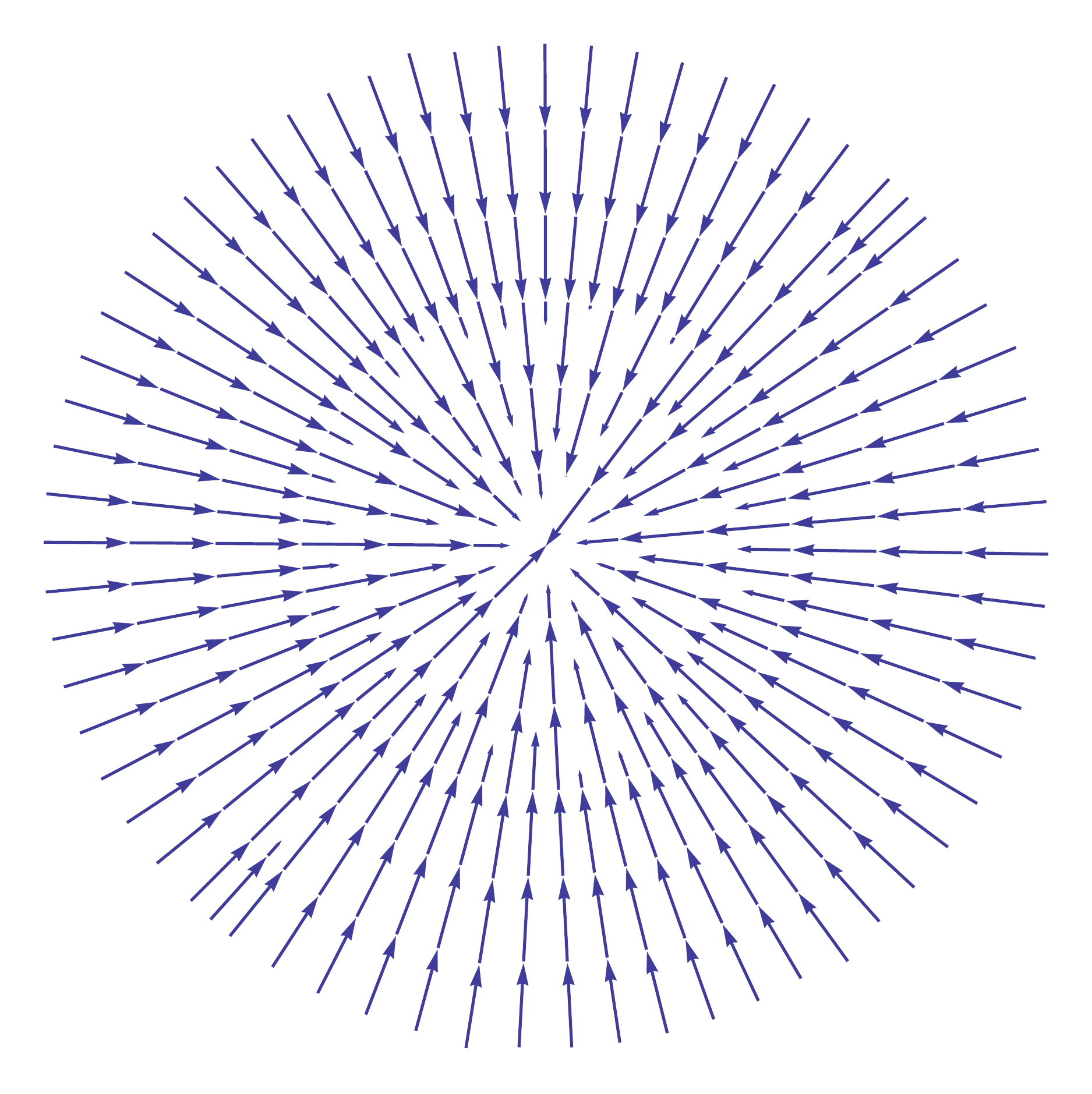}
\caption{\small Symmetric $\theta$-independent flow of the inflaton field to the minimum of the potential in the model (\ref{2dt}) shown in Fig. \ref{tmodelfig1}.}
\label{tmodelfig1ss}
\end{figure}

If the potential does not depend on $\theta$, like the potential (\ref{2dt}), then, according to the slow-roll equation for $\theta$, one has $\dot \theta = 0$, so this field does not move, and the field $\vp$ obeys the standard single-field slow roll equation
\be
3H\dot \vp =  - {dV\over d\vp}\ .
\ee
The flow of the fields during inflation in this case is described by the spherically symmetric distribution shown in Figure \ref{tmodelfig1ss}. 

This means that in spherically symmetric potentials, the field evolution is effectively one-dimensional, and equations describing it and its cosmological consequences are the same as in the single-field case described in section \ref{single}. The only difference is related to the fluctuations of the light field $\theta$ which may be generated during inflation. These perturbations typically are harmless (and useless), unless one considers some special versions of the theory which make post-inflationary decay of the energy of the field $\theta$ extremely strongly suppressed. In that case, they may serve the role of isocurvature perturbations \cite{Linde:1984ti}, or may become responsible for the curvaton mechanism of generation of adiabatic perturbations \cite{curva,Demozzi:2010aj}. Another possibility is to consider theories where reheating is modulated by the fluctuations of the field $\theta$ \cite{Dvali:2003em}. In what follows, we will not consider these possibilities.

Under this condition, all results concerning naturalness of inflation and universality of predictions for $n_{s}$ and $r$ in the class of single-field models considered in section  \ref{single} remain valid for arbitrary spherically symmetric multi-field potentials of the type considered above.

\begin{figure}[ht!]
\centering
\includegraphics[scale=0.28]{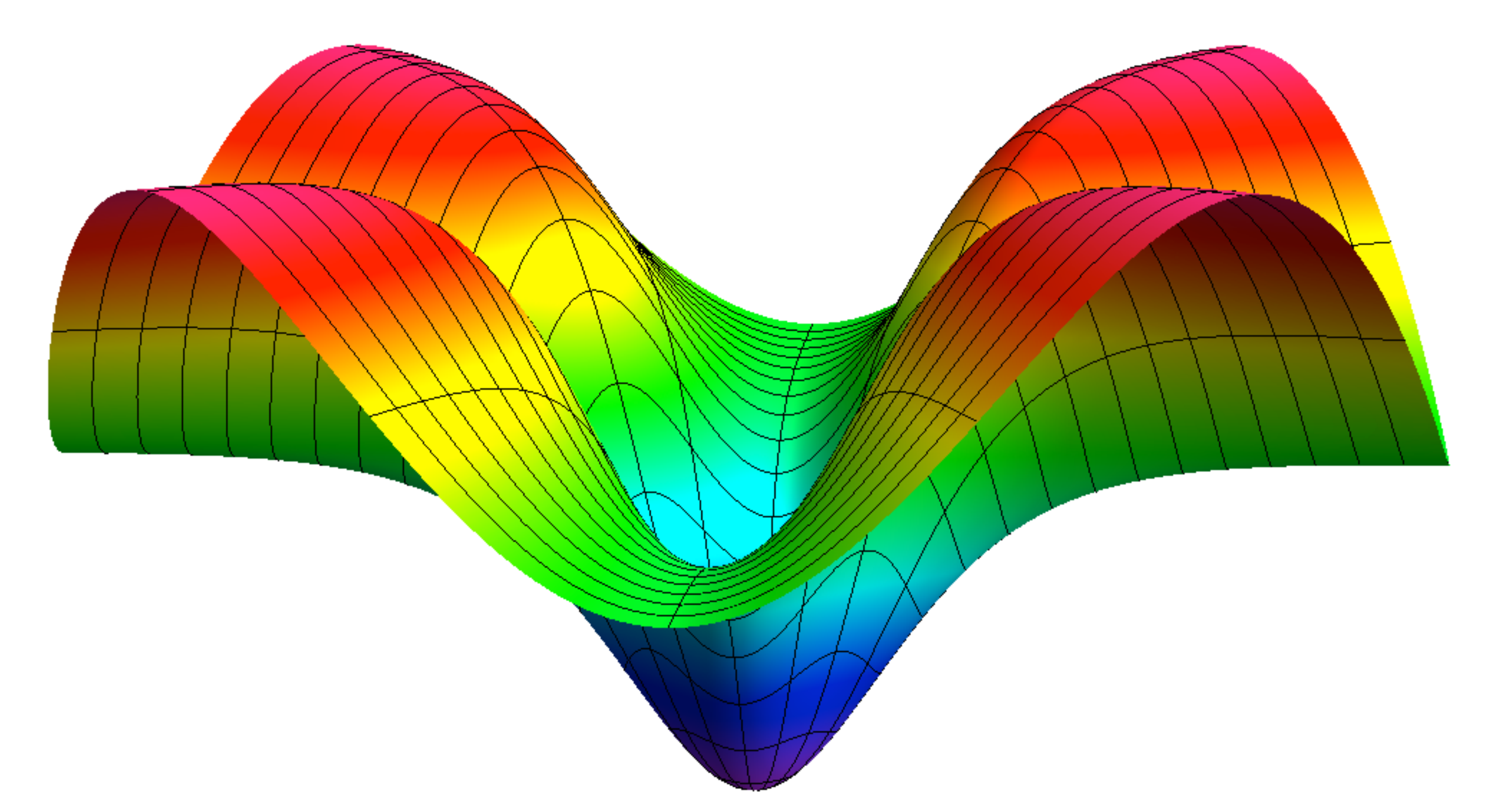}
\caption{\small Potential (\ref{flower}) for $A = B$.}
\label{tmodelfig2}
\end{figure}

Now we will turn to more complicated potentials depending both on $\vp$ and on $\theta$, and analyze the field evolution in the slow roll approximation at large $\vp$, where $\sinh{\vp\over\sqrt 6} \approx {1\over 2} e^{\vp/\sqrt 6}$.  As a first example, we will consider the following simple but instructive model: 
\be
F\left(z_{i}\right) = A\, (z_{1}^{2} + z^{2}_{2}) +4B\, z^{2}_{1}z^{2}_{2}
\ee
where $z_{i} = {\phi_{i}\over \xi}$. The resulting Einstein frame potential is 
\be\label{flower}
V = A\, {\tanh^{2}{\varphi\over \sqrt 6}}+ {B}\, {\tanh^{4}{\varphi\over \sqrt 6}}\ \sin^{2} 2\theta
\ee
This function is shown in Fig. \ref{tmodelfig2} for $A = B$.

At large $\vp$, equations for the fields $\theta$ and $\vp$ in this model look as follows:
\be
3H\dot\theta = -{4B\, \sin {4\theta}\over 3} \  e^{-\sqrt{2\over 3}\vp}\ ,
\ee
\be
3H\dot \vp = - 4\sqrt{2\over 3}\ \left(A +2B \sin^{2} 2\theta- {{B}^{2} \sin^{2} 4\theta \over 9(A+{B} \sin^{2} 2\theta)}\, \right)\,  e^{-\sqrt{2\over 3}\vp} \ .
\ee

These equations imply the following equation relating the behavior of $\theta$ during the rolling  of the field $\vp$ to the minimum of the potential at $\vp = 0$:
\be\label{tp}
{d\theta\over d\vp} ={ {B\, \sin {4\theta}} \over {\sqrt{6}} \left(A +2B \sin^{2} 2\theta - {{B}^{2} \sin^{2} 4\theta \over 9(A+{B} \sin^{2} 2\theta)}\right)}\ .
\ee
\begin{figure}[ht!]
\centering
\includegraphics[scale=0.41]{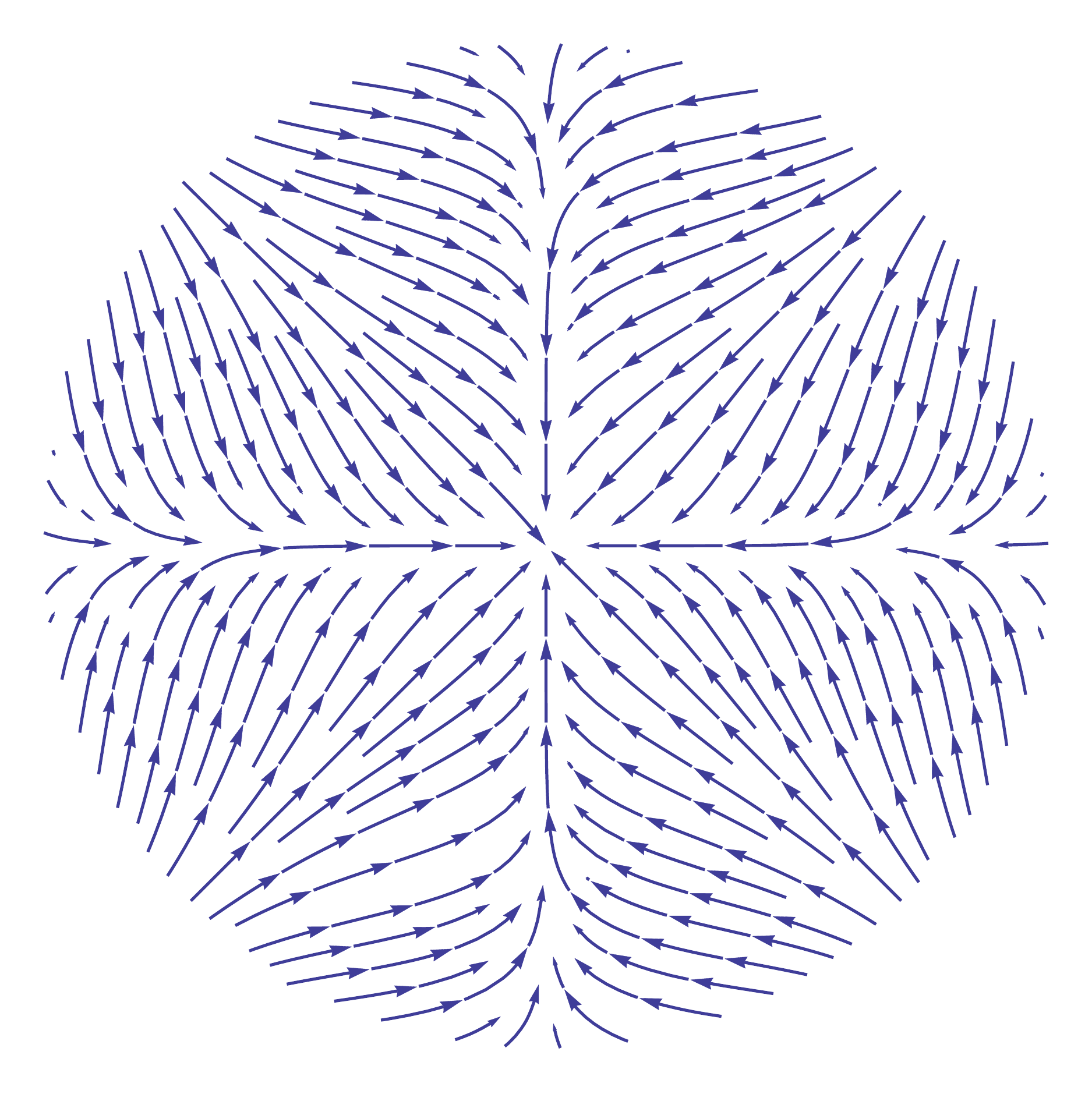}
\caption{\small Asymmetric $\theta$-dependent flow of the fields $\vp$ and $\theta$ to the minimum of the potential in the model (\ref{flower}) for $A = B$. The flow is shown at late stages of inflation, starting at $\vp_{0} = 8$. This initial boundary corresponds to the circular boundary of the region shown in this figure. }
\label{tmodelfig12s}
\end{figure}

\begin{figure}[ht!]
\centering
\hskip 0.6cm \includegraphics[scale=0.32]{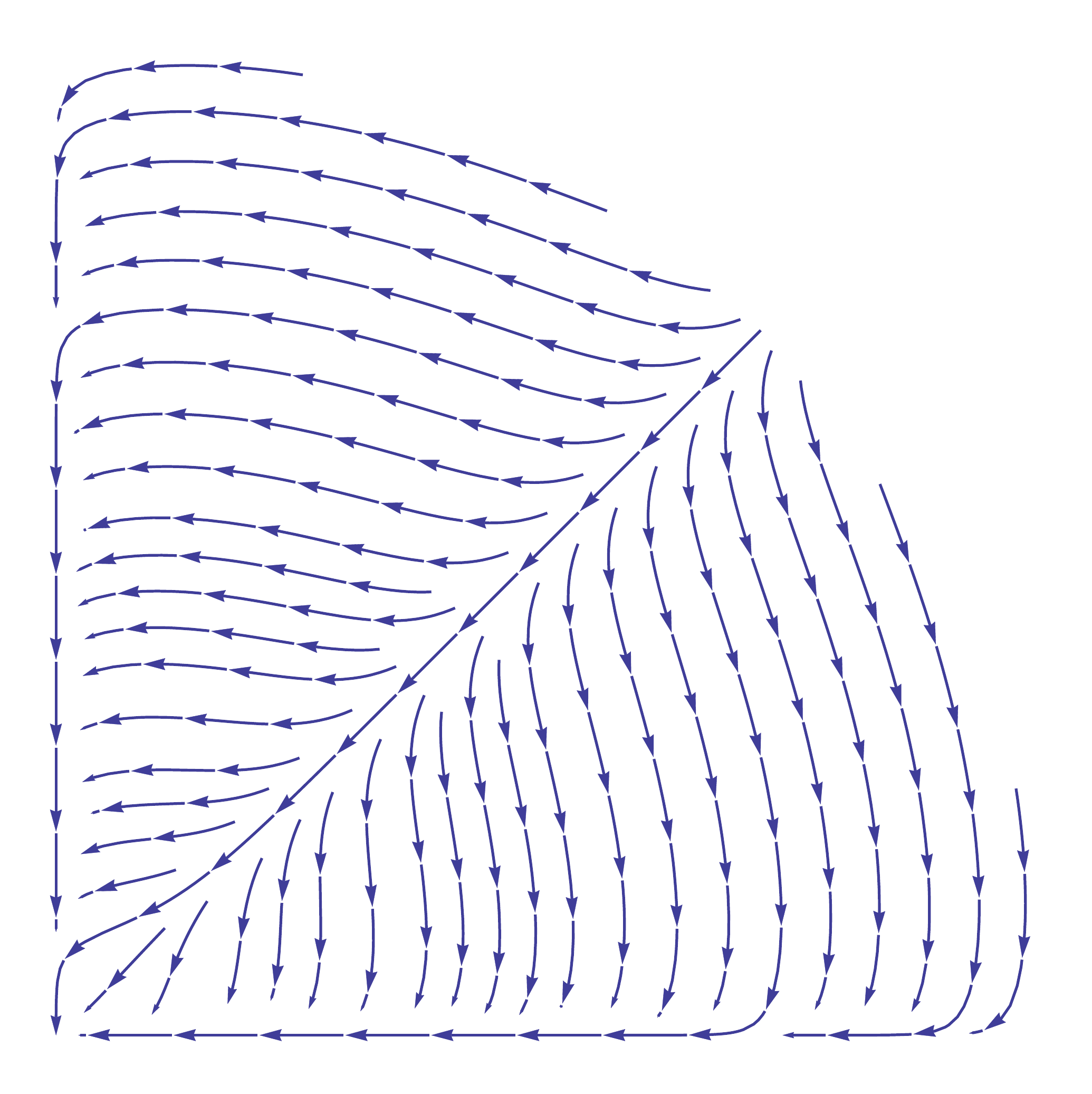}
\caption{\small The flow of the fields $\vp$ and $\theta$ to the minimum of the potential in the model (\ref{flower}) for $A = B$ in the quadrant $0<\theta < \pi/2$. The flow is shown at earlier stages of inflationary evolution, starting at $\vp_{0} = 40$. As we see, comparing it to Figure \ref{tmodelfig12s}, at large $\vp$ the trajectories reach the valleys faster, in terms of the relative change $d\vp/\vp$.}
\label{tmodelfig12ss}
\end{figure}

One can plot the trajectory $\theta(\vp)$ in polar coordinates, by decomposing the motion to its radial component $\dot\vp$ and its angular component $\vp\,\dot\theta$ as shown in Figure \ref{tmodelfig12s}. In this figure we show the evolution starting at $\vp_{0} = 8$, to give a particular example. As we see, the flow of inflationary trajectories takes the fields away from the ridges of the potential at $\theta = \pi/4 + n\pi/2$ and brings it very close to the valleys at $\theta = n\pi/2$. 
One can see it even better if one draws the set of inflationary trajectories starting at much greater values of the field $\vp_{0}$, see Figure \ref{tmodelfig12ss}. This figure shows the flow of the fields in one quadrant, $0<\theta < \pi/2$, starting at $\vp_{0} = 40$, for the same values $A = B$ as in the previous figure  \ref{tmodelfig12s}. As we see, at large $\vp$ the field trajectories move almost orthogonally to the ridge of the potential at $\theta = \pi/4$ and quickly reach the valleys at $\theta = \pi/2$ and at $\theta = 0$. The reason is that the speed of motion of both fields $\theta$ and $\vp$ is suppressed by the same exponent $e^{-\sqrt{2/3}\, \vp}$, but the range of the evolution of the field $\theta$ is $\Delta \theta = O(1)$, whereas the evolution of the field $\phi$ may start at its indefinitely large values. Note that ${d\theta\over d\vp}$ does not depend on $\vp$, which means that at large $\vp$ the evolution of the field $\theta$ occurs faster field  in terms of the relative change of $d\vp/\vp$.

An investigation of the evolution of the fields $\theta$ and $\vp$ is especially transparent if one considers the limit $\theta \ll 1$ near the valley at $\theta = 0$. In this case equation (\ref{tp}) acquires a simple form
\be
{d\theta\over d\vp} =-{ 4B \over {\sqrt{6}} A}\,  \theta \ ,
\ee
which has a solution $\theta = \theta_{0}\ \exp\bigl({4B\over \sqrt 6 A}(\vp-\vp_{0})\bigr)$.
For $\theta_{0}\lesssim 1$, one finds  that if the slow roll evolution begins at $\vp_{0} \gg 1$, then in the end of inflation, where $\vp \ll \vp_{0}$, one has
\be\label{thetaphi}
\theta <  e^{-{4B\over \sqrt 6 A}\vp_{0}} \ . 
\ee
In other words, for any reasonable initial conditions at $\vp_{0} \gg 1$, the field $\theta$ during the last 60 e-foldings stays exponentially close to the bottom of the valley (i.e. to the minimum of the potential in the $\theta$ direction), and therefore it does not participate in inflationary evolution. This means that the field evolution becomes effectively one-dimensional. As a result, all qualitative conclusions of Ref. \cite{Kallosh:2013hoa} concerning universality of cosmological predictions of his class of models remain applicable to the theories with many scalars.

Some additional comments are in order here. First of all, the speed of convergence to the single-field regime in this class of models depends on the parameters of the potential. For example, for $A \gg B$, this speed may be very limited, so the convergence occurs only if one can consider extremely large values of $\vp_{0}$. Our expectations of what could be $\vp_{0}$ may depend on the way these theories are embedded in the string theory, or on the process of quantum creation of the universe with large $\vp_{0}$. If we take an agnostic position here, we may say that more complicated regimes with combined evolution of $\vp$ and $\theta$ depicted in Figure \ref{tmodelfig12s} are also possible, which may lead to more complicated spectra of cosmological perturbations. On the other hand, in the limit $A \gg B$, the flow of the fields down to the minimum of the potential during the last 60 e-foldings becomes nearly exactly spherically symmetric, indistinguishable from what we see in Figure \ref{tmodelfig1ss}, in which case the classical evolution of the field $\theta$ becomes irrelevant anyway.

\begin{figure}[ht!]
\centering
\includegraphics[scale=0.22]{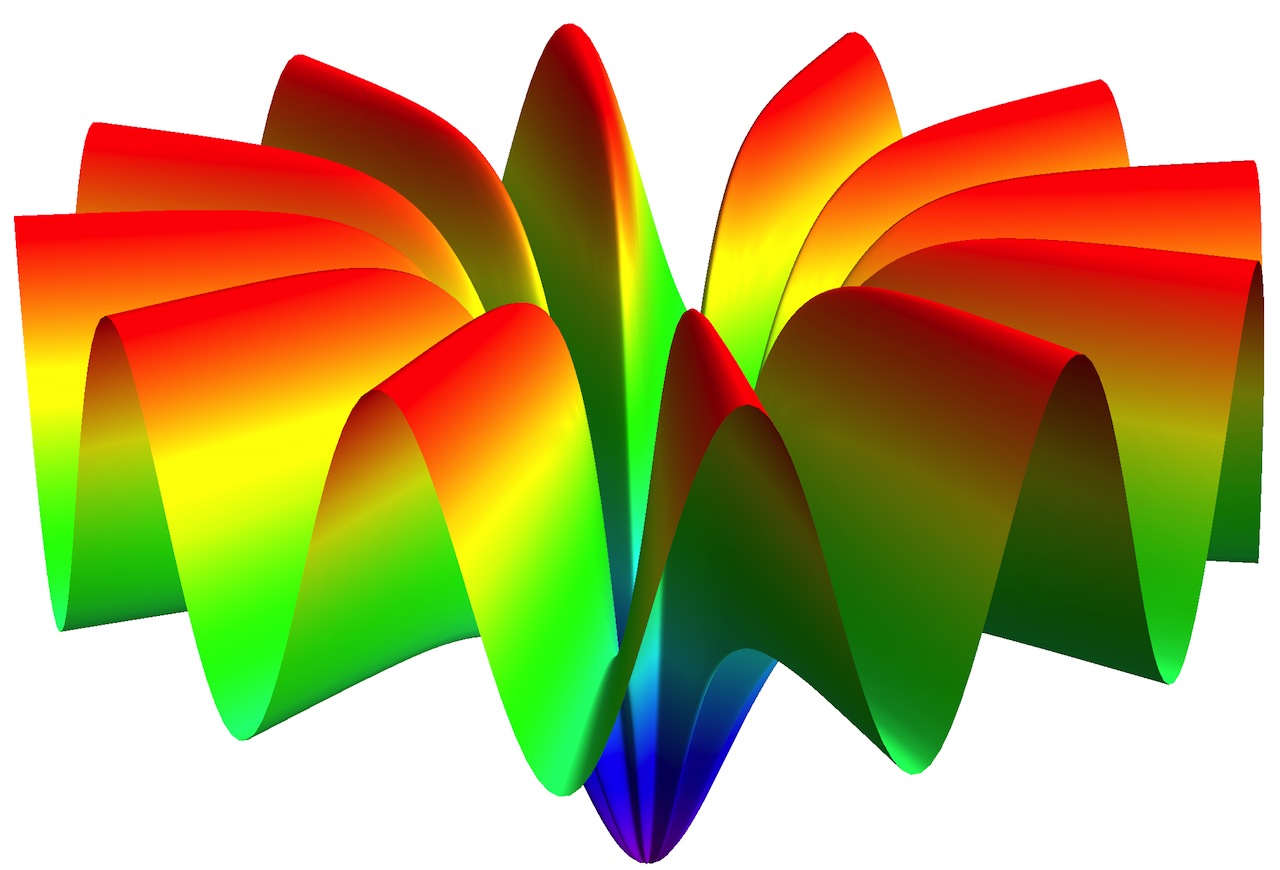}
\caption{\small Potential $V = {\tanh^{2}{\varphi\over \sqrt 6}} (A + {B}\  \sin^{2} 2n\theta)$ for $A = B$, $n = 3$.}
\label{tmodelfig2a}
\end{figure}

In the opposite limit, $A \ll B$, the convergence to the single-field regime occurs very fast. One can achieve the same result by increasing the curvature of the potential with respect to $\theta$. For example, one may consider the Einstein frame potential 
\be
V = {\tanh^{2}{\varphi\over \sqrt 6}} (A + {B}\  \sin^{2} 2n\theta)
\ee
This potential is shown in Figure \ref{tmodelfig2a} for the case $A = B$, $n = 3$.
This change will increase the factor in the exponent in (\ref{thetaphi}) $n$ times, which will lead to a much faster relaxation of the field $\theta$ towards one of the valleys of the potential and the classical field evolution becomes effectively one-dimensional.


However, one should note that unless one takes extremely large values of $n$ or tune other parameters of the potential, the effective mass squared of all scalar fields, including the field $\theta$, typically remain smaller than $H$ during the main part of the inflationary period. In order to check it, one may either evaluate the mass matrix of all fields, or simply look at the equations of motion for the fields $\theta$ and $\vp$ and find that in both cases, the steepness of the effective potential of each of these fields (taking into account nonminimal kinetic term of the field $\theta$) is suppressed by the same exponential factor $e^{-\sqrt{2\over 3}\vp}$. Thus the lightness of the inflaton field mass ($|m_{\vp}^{2}| \ll H^{2}$ during inflation) typically goes hand in hand with the lightness of the field $\theta$.

As we already mentioned, quantum fluctuations orthogonal to the inflationary trajectory typically do not lead to the problematic isocurvature fluctuations unless one makes some special arrangements to extremely strongly suppress the decay of the field $\theta$ at the end of inflation. On the other hand,  the existence of more than one light scalars in the multi-field models of conformal inflation may make this group of theories a natural starting point in a search for additional sources of inflationary perturbations, based e.g. on the curvaton scenario \cite{curva,Demozzi:2010aj} or on the theory of modulated preheating \cite{Dvali:2003em}.

\section{Generic potentials of the multi-field conformal inflation}\label{TT}

\begin{figure}[ht!]
\centering
\includegraphics[scale=0.2]{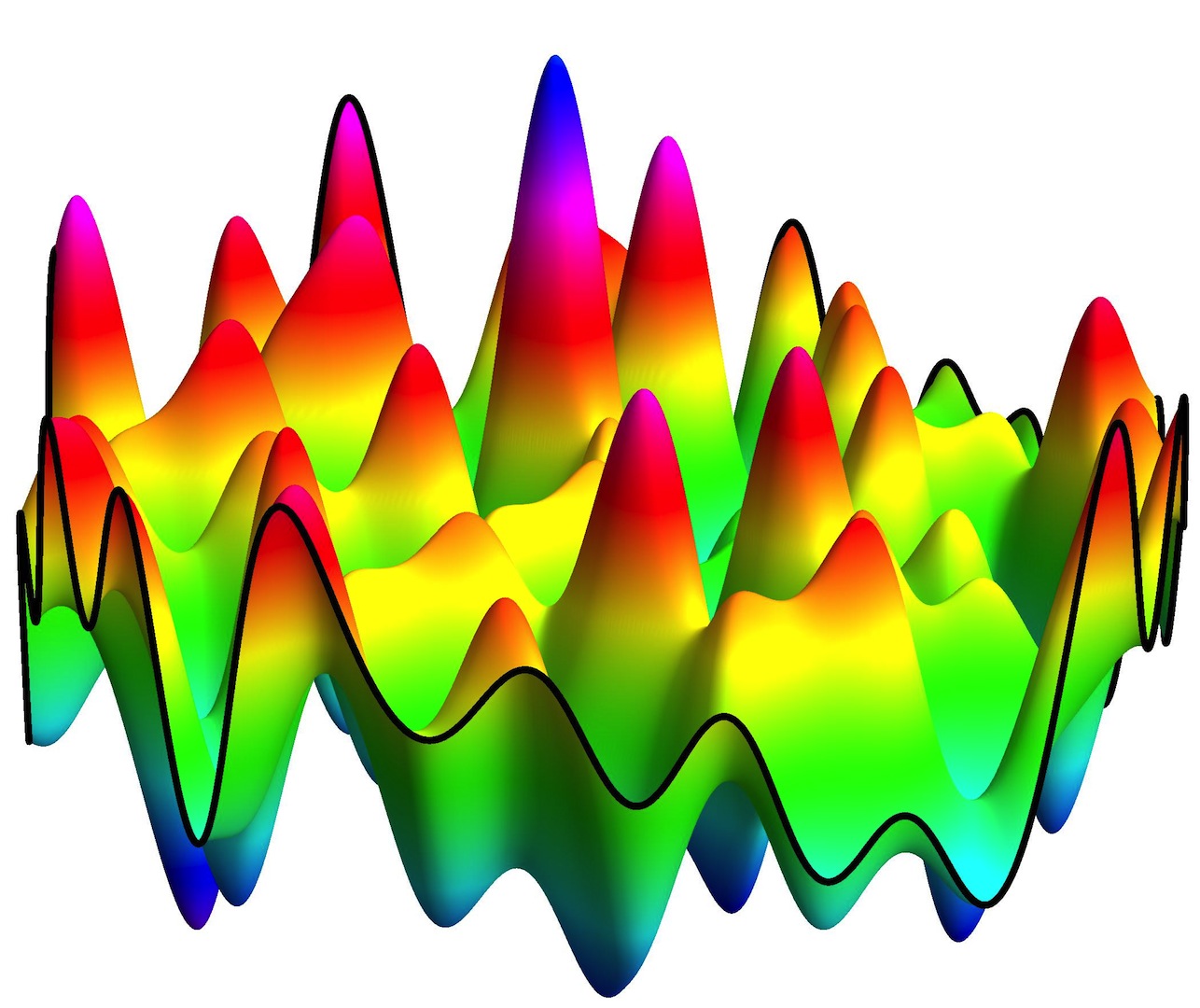}
\caption{\small Random function $F\left(z_{i}\right)$ in terms of the conformal variables $z_{i} = {\phi_{i}\over \chi}$. The black  line shows the boundary of the moduli space $|z_{i}|^{2} = 1$. This boundary moves to infinity upon switching to the canonically normalized radial variable $\vp$ in the Einstein frame. This is the main reason of the stretching of the potential in the radial direction, as shown in Figures \ref{stretch} and \ref{tmodelfig4}.}
\label{tmodelfig3}
\end{figure}

In the previous section we studied a special class of models which had a very simple dependence on $\rho$ and $\theta$. Now we are prepared for a discussion of a more general case. Along the lines of Section \ref{single}, we will consider, as an example, a very chaotically looking function $F\left(z_{i}\right)$ shown in Figure \ref{tmodelfig3}, which is a generalization of a chaotic distribution shown in the upper panel of Figure \ref{stretch}. We generated this potential by the same method that one can use for simulations of random Gaussian fluctuations with a flat spectrum, with a UV cut in the spectrum corresponding to the smallest inhomogeneities visually identifiable there. The boundary of the moduli space in terms of the variables $z_{i}$ is at $z_{i}^{2} = 1$, which means that the corresponding radius is $|z| = {\rho\over \xi} = 1$.

\begin{figure}[ht!]
\centering
\includegraphics[scale=0.32]{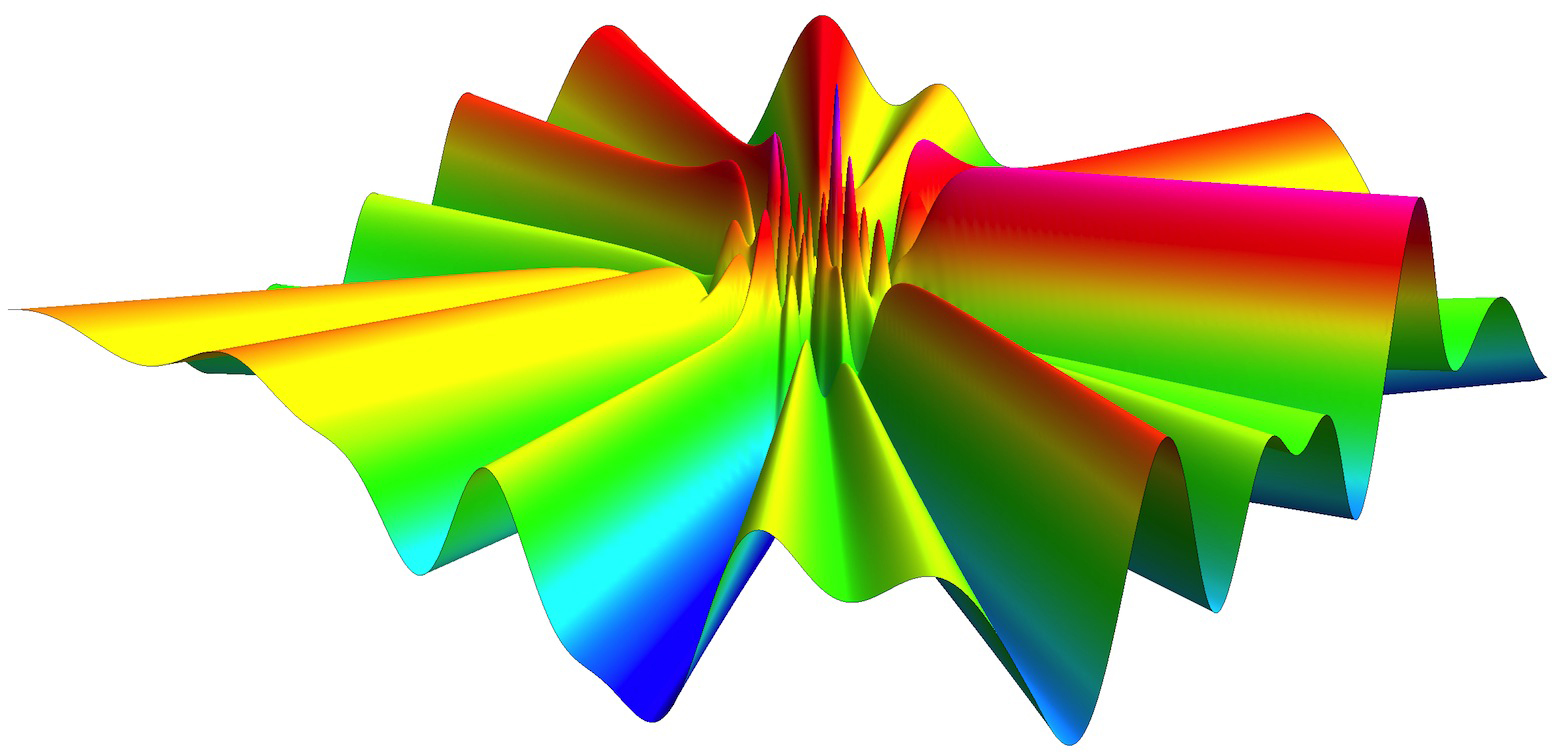}
\caption{\small The Einstein frame potential corresponding to the random function $F\left(z_{i}\right)$ shown in Figure \ref{tmodelfig3}. Yellow and light green valleys are inflationary directions with $V> 0$, blue valleys correspond to $V<0$.}
\label{tmodelfig4}
\end{figure}

The potential looks very steep and disorderly, it does not contain any nicely looking flat regions, so naively one would not expect it to lead to a slow-roll inflation. Indeed, upon switching to the canonically normalized field $\vp$ in the Einstein frame, the central part of this distribution stretches by the factor of $\sqrt 6$, but it remains chaotic and mostly unsuitable for inflation. However, just as in the one-field case, the boundary of the moduli space experiences an infinitely large stretching, forming infinitely long and straight ridges and valleys. What looked like sharp Minkowski or dS minima near the boundary become the final destinations for inflationary evolution of the field rolling towards smaller values of $\vp$ along de Sitter valleys, see Figure \ref{tmodelfig4}. (That is where the color coding is especially important: The yellow or light green colors do not change along the valleys; they change only close to the center of the figure where the field rolls down to the minima of the potential.) Meanwhile the blue valleys correspond to negative cosmological constant; we do not want to go there because the corresponding parts of the universe rapidly collapse.

As we already mentioned, our figures, including Figure \ref{tmodelfig4}, under-represent the angular stretching of the moduli space: The true distance between the different valleys is proportional not to the radial distance $\vp$, as the figures might suggest, but to $\sqrt 6 \sinh {\varphi\over \sqrt 6}$. That is why the ridges and valleys are in fact exponentially soft in all directions, and the fields can eternally wonder in these mountains due to inflationary quantum fluctuations, unless some of them fall to AdS minima.

Thus what we have is a partially traversable landscape, divided into semi-infinite inflationary areas separated by the deep blue valleys with negative vacuum energy, corresponding to collapsing parts of the multiverse.
One may conjecture that in the theories involving more than 2 different scalars, having many different inflationary valleys, the percolation between different inflationary valleys become possible despite the existence of the valleys with negative values of the potential.

The structure of the inflationary valleys is determined by the properties of the function $F\left(z_{i}\right)$ in the close vicinity of the boundary of the moduli space. The sharper are the minima of the function $F\left(z_{i}\right)$, the more of these minima fit near the boundary, the greater is the variety of inflationary valleys we are going to obtain. If a typical size of such sharp minima is equal to $\Delta z$, the number of different inflationary valleys slowly bending towards different Minkowski or dS minima should be proportional to $(\Delta z)^{n-1}$, where $n$ is the total number of the different moduli. For $\Delta z \ll 1$ and $n\gg 1$, one can get an exponentially large variety of different possibilities, reminiscent of the string theory landscape.

It is instructive to compare this scenario with the more conventional multi-field scenario. If one assumes that from the very beginning we deal with a random Einstein frame potential without any symmetries protecting its flat directions, one may conclude that such flat directions are rather unlikely, see e.g. \cite{Marsh:2013qca}. On the other hand, the more chaotic is the function $F$ in terms of the original conformal variables $z_{i}$, and the greater number such fields we have, the greater is the variety of inflationary valleys which naturally and nearly unavoidably emerge in the scenario outlined in our paper. 

Now let us turn to the predictions of this class of theories. As we already mentioned in the previous section, barring some fine-tuning of initial conditions and parameters of the theory, the field $\theta$ usually has plenty of time to roll down to one of the inflationary valleys, and after that the classical evolution is entirely determined by the single-field evolution of the field $\vp$. 
The main reason is that the speed of motion of both fields $\theta$ and $\vp$ is suppressed by the same exponent $e^{-\sqrt{2/3}\, \vp}$, but the range of the evolution of the field $\theta$ is $\Delta \theta = O(1)$, whereas the evolution of the field $\phi$ may start at its indefinitely large values, as one may conclude by looking at Figure \ref{tmodelfig4}. As a result, the field $\theta$ typically rolls down exponentially closely to the minimum of its potential long before the last 60 e-foldings of inflation, see (\ref{thetaphi}), and the subsequent evolution is determined by the field $\vp$. This leads to universal observational predictions $1 -n_{s} =2/N$, $r = 12/N^{2}$, just as for the single-field attractors studied in \cite{Kallosh:2013hoa}.

\section{Conclusions}

In this paper we developed a multi-field generalization of the new class of conformal and superconformal inflationary models with the universal attractor behavior  \cite{Kallosh:2013hoa}. Our approach can be easily extended to multi-field models with a general negative non-conformal coupling to gravity $\xi < 0$  \cite{Kallosh:2013maa}. We hope to return to this issue, and also to study of superconformal generalizations of these models. Since the basic reason for the existence of the cosmological attractors described above (stretching of the boundary of the moduli space) is quite generic, we expect that many new examples of this mechanism are waiting to be discovered.

\subsection*{Acknowledgments}

We are  grateful to D. Kaiser, V. Mukhanov, E. Sfakianakis, E. Silverstein and L. Susskind for stimulating discussions. 
This work  is supported by the SITP and by the 
NSF Grant No. 0756174.

\end{document}